\documentclass[prl,twocolumn,showpacs,floatfix]{revtex4}
\usepackage{amsmath,amssymb}
\usepackage{graphicx}
\usepackage{bbm}
\usepackage{dcolumn}

\newcommand{\1}{\mathbbm{1}}
\newcommand{\eps}{\varepsilon}
\newcommand{\ev}[1]{\left\langle#1\right\rangle}
\DeclareMathOperator{\diag}{diag}
\DeclareMathOperator{\re}{Re}
\DeclareMathOperator{\sign}{sign}
\DeclareMathOperator{\tr}{tr}
\DeclareMathOperator{\ind}{index}

\begin{document}

\title{Overlap Dirac operator at nonzero chemical potential and random
  matrix theory}

\author{Jacques Bloch and Tilo Wettig}
\affiliation{Institute for Theoretical Physics, University of
  Regensburg, 93040 Regensburg, Germany}

\date{April 24, 2006}

\begin{abstract}
  We show how to introduce a quark chemical potential in the overlap
  Dirac operator.  The resulting operator satisfies a Ginsparg-Wilson
  relation and has exact zero modes.  It is no longer
  $\gamma_5$-hermitian, but its nonreal eigenvalues still occur in
  pairs.  We compute the spectral density of the operator on the
  lattice and show that, for small eigenvalues, the data agree with
  analytical predictions of nonhermitian chiral random matrix theory
  for both trivial and nontrivial topology.
\end{abstract}



\maketitle

Recent years have seen great advances in two areas that at first sight
seem to be totally unrelated: (i) the study of nonhermitian operators
in the natural sciences \cite{pseudospectra} and (ii) the problem
of chiral symmetry on the lattice \cite{Ha04}.  In this article we
focus on a problem in which both of these areas are relevant,
namely quantum chromodynamics (QCD) at nonzero baryon (or quark)
density, which is important for the study of relativistic heavy-ion
collisions, neutron stars, and the early universe \cite{owe}.

If a quark chemical potential $\mu$ is added to the QCD Dirac
operator, the operator loses its hermiticity properties and its
spectrum moves into the complex plane.  This causes a variety of
problems, both analytically and numerically.  Lattice simulations are
the main source of nonperturbative information about QCD, but at
$\mu\ne0$ they cannot be performed by standard importance sampling
methods because the measure of the Feynman path integral, which
includes the complex fermion determinant, is no longer positive
definite.  While a generic solution to this so-called sign problem is
unlikely to be found \cite{tw05}, a number of recent works have been
able to make progress by circumventing the problem in various ways
\cite{FK02,BielSwan02,dFP02}.  These methods all agree on the
transition temperature from the hadronic to the quark-gluon phase in
the regime $\mu/T\lesssim1$ \cite{owe}.

A better analytical understanding of QCD at very high baryon density
has been obtained by a number of methods \cite{Al02}, and the QCD
phase diagram has been studied in model calculations based on
symmetries \cite{HJSSV98}.  Chiral random matrix theory (RMT)
\cite{SV93}, which makes exact analytical predictions for the
correlations of the small Dirac eigenvalues, has been extended to
$\mu\ne0$ \cite{Step96}, and a mechanism was identified \cite{OSV05}
by which the chiral condensate at $\mu\ne0$ is built up from the
spectral density of the Dirac operator in an extended region of the
complex plane, in stark contrast to the Banks-Casher mechanism at
$\mu=0$.

A first comparison of lattice data with RMT predictions at $\mu\ne0$
was made in Ref.~\cite{AW04} using staggered fermions.  One issue with
staggered fermions is that the topology of the gauge field is only
visible in the Dirac spectrum if the lattice spacing is small and
various improvement and/or smearing schemes are applied
\cite{stag-top}.  To avoid these issues, we would like to work with a
Dirac operator that implements a lattice version of chiral symmetry
and has exact zero modes at finite lattice spacing.  The overlap
operator \cite{overlap} satisfies these requirements at $\mu=0$.  In
the following, we show how the overlap operator can be modified to
include a nonzero quark chemical potential \cite{BW98,BH00}.  We then
study the spectral properties of this operator as a function of $\mu$
and compare data from lattice simulations with RMT predictions.  As we
shall see, the overlap operator has exact zero modes also at nonzero
$\mu$, which allows us, for the first time, to test predictions of
nonhermitian RMT for nontrivial topology.

We begin with the well-known definition of the Wilson Dirac operator
$D_W$ including a chemical potential $\mu$ \cite{hk83},
\begin{align}
  D_W(\mu)&=\1-\kappa\sum_{i=1}^3\left(T_i^++T_i^-\right)
  -\kappa\left(e^\mu T_4^++e^{-\mu} T_4^-\right)\:,\notag\\
  (T_\nu^\pm)_{yx}&=(1\pm\gamma_\nu)U_{\pm\nu}(x)\delta_{y,x\pm\hat\nu}\:,
  \label{eq:dwmu}
\end{align}
where $\kappa=1/(2m_W+8)$ with the Wilson mass $m_W$, the
$U\in\text{SU(3)}$ are the lattice gauge fields, and the $\gamma_\nu$
are the usual euclidean Dirac matrices.  Unless displayed explicitly,
the lattice spacing $a$ is set to unity.  The overlap operator is
defined at $\mu=0$ by \cite{overlap}
\begin{align}
  \label{eq:overlap0}
  D_\text{ov}(0)=\1+\gamma_5\eps(\gamma_5D_W(0))\:,
\end{align}
where $\eps$ is the matrix sign function and
$\gamma_5=\gamma_1\gamma_2\gamma_3\gamma_4$.  $m_W$ must be in the
range $(-2,0)$ for $D_\text{ov}(0)$ to describe a single Dirac fermion
in the continuum.  The properties of $D_\text{ov}(0)$ have been
studied in great detail in the past years.  In particular, its
eigenvalues are on a circle in the complex plane with center at
$(1,0)$ and radius 1, its nonreal eigenvalues come in complex
conjugate pairs, and it can have exact zero modes without fine-tuning.
$D_\text{ov}(0)$ satisfies a Ginsparg-Wilson relation \cite{GW82} of
the form
\begin{align}
  \label{eq:gw}
  \{D,\gamma_5\}=D\gamma_5 D\:.
\end{align}

We now extend the definition of the overlap operator to $\mu\ne0$.
The operator $D_W(0)$ in Eq.~\eqref{eq:overlap0} is
$\gamma_5$-hermitian, i.e., $\gamma_5D_W(0)\gamma_5=D_W^\dagger(0)$,
and therefore the operator $\gamma_5D_W(0)$ in the matrix sign
function is hermitian.  However, for $\mu\ne0$, $D_W(\mu)$ is no
longer $\gamma_5$-hermitian.  Defining the overlap operator at nonzero
$\mu$ by
\begin{align}
  \label{eq:overlapmu}
  D_\text{ov}(\mu)=\1+\gamma_5\eps(\gamma_5D_W(\mu))\:,
\end{align}
we now need the sign function of a nonhermitian matrix.  In general, a
function $f$ of a nonhermitian matrix $A$ can be defined by a contour
integral.  A more convenient expression can be obtained if $A$ is
diagonalizable.  In this case we can write $A=U\Lambda U^{-1}$, where
$U\in\text{Gl}(N,\mathbb{C})$ with $N=\dim(A)$ and
$\Lambda=\diag(\lambda_1,\ldots,\lambda_N)$ with
$\lambda_i\in\mathbb{C}$.  Then $f(A)=Uf(\Lambda)U^{-1}$, where
$f(\Lambda)$ is a diagonal matrix with elements $f(\lambda_i)$.  The
sign function can be defined by \cite{sign}
\begin{align}
  \label{eq:sign}
  \eps(A)=U\sign(\re\Lambda)U^{-1}\:.
\end{align}
This definition ensures that $\eps^2(A)=\1$ and gives the correct
result if $\Lambda$ is real.  An equivalent definition is
$\eps(A)=A(A^2)^{-1/2}$ \cite{H94}.  Eqs.~\eqref{eq:overlapmu} and
\eqref{eq:sign} constitute our definition of $D_\text{ov}(\mu)$.  The
sign function is ill-defined if one of the $\lambda_i$ lies on the
imaginary axis.  Also, it could happen that $\gamma_5D_W(\mu)$ is not
diagonalizable (one would then resort to a Jordan block
decomposition).  Both of these cases are only realized if one or more
parameters are fine-tuned.  This is unlikely to happen in realistic
lattice simulations, and we therefore ignore these issues.

It is relatively straightforward to derive the following properties of
$D_\text{ov}(\mu)$.  \leftmargini3mm
\begin{itemize}\itemsep-1.3mm
\item $D_\text{ov}(\mu)$ is no longer $\gamma_5$-hermitian because for
  a nonhermitian matrix $A$ we generically have
  $\eps^\dagger(A)\ne\eps(A)$.  Instead, $D_\text{ov}(\mu)$ satisfies
  $\gamma_5D_\text{ov}(\mu)\gamma_5=D_\text{ov}^\dagger(-\mu)$.  
\item $D_\text{ov}(\mu)$ still satisfies the Ginsparg-Wilson relation
  \eqref{eq:gw} because of $\eps^2(A)=\1$.  Thus, we still have a
  lattice version of chiral symmetry, and the operator has exact zero
  modes without fine-tuning.  
\item The eigenvalues of $D_\text{ov}(\mu)$ that are not equal to 0
  or 2 no longer come in complex conjugate pairs, but every such
  eigenvalue $\lambda$ (with eigenfunction $\psi$) is accompanied by a
  partner $\lambda/(\lambda-1)$ (with eigenfunction $\gamma_5\psi$).
\item At $\mu=0$, the mapping $\lambda\to z=2\lambda/(2-\lambda)$
  projects the eigenvalues from the circle to the imaginary axis.  At
  $\mu\ne0$, the same mapping projects the eigenvalues $\lambda$ and
  $\lambda/(\lambda-1)$ to a pair $\pm z$, which is the same pairing
  as for $\mu\ne0$ in the continuum.
\item The eigenfunctions of $D_\text{ov}(\mu)$ corresponding to
  eigenvalue 0 or 2 can be arranged to have definite chirality.  The
  numbers $n_\lambda^\pm$ of modes with $\lambda=0,2$ and
  $\ev{\gamma_5}=\pm1$ satisfy $n_0^+-n_0^-=-(n_2^+-n_2^-)$.  (Without
  fine-tuning, we even have $n_0^-=n_2^+=0$ or $n_0^+=n_2^-=0$, which
  is always the case in our lattice data below.)
\item Generically, $D_\text{ov}(\mu)$ is not normal, i.e.,
  $DD^\dagger\ne D^\dagger D$.
\end{itemize}

We now turn to our (quenched) lattice simulations.  The computation of
the sign function of a nonhermitian matrix is very demanding.  We are
currently investigating various approximation schemes \cite{bflw}, but
in this initial study we decided to compute the sign function and to
diagonalize $D_\text{ov}(\mu)$ exactly using LAPACK.  For the
comparison with RMT we need high statistics, which restricts us to a
very small lattice size.  We have chosen the same parameter set as in
Ref.~\cite{ehkn99} to be able to compare with previous results at
$\mu=0$.  The lattice size is $V=4^4$, the coupling in the standard
Wilson action is $\beta=5.1$, the Wilson mass is $m_W=-2$, and the
quark mass is $m_q=0$.  The small lattice size forces us to use such
a strong coupling to stay in the ergodic regime of QCD, see
Eq.~\eqref{eq:ergodic} below, but we expect the conclusions of this
paper to remain valid at weaker couplings and correspondingly larger
lattice sizes.  The choice of $m_W=-2$ is motivated by concerns about
the locality of $D_\text{ov}(\mu)$, although at $V=4^4$ this question
is largely academic.  

\begin{figure}[-t]
  \centerline{\includegraphics[height=35mm]{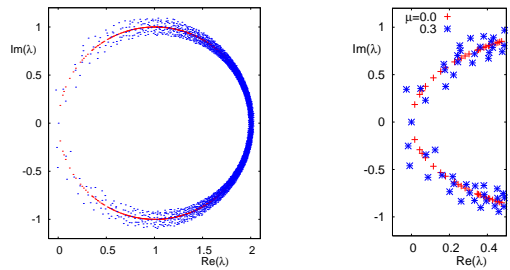}}
  \caption{Spectrum of $D_\text{ov}(\mu)$ for $\mu=0$ and $\mu=0.3$
    for a typical configuration.  The figure on the right is a
    magnification of the region near zero.}
  \label{fig:spectrum}
\end{figure}

\begin{table}[-b]
  \caption{Number of configurations and distribution of the number
    $\nu$ of zero modes  of $D_\text{ov}(\mu)$ for various values of $\mu$.}
  \label{table1}
  \begin{ruledtabular}
    \begin{tabular}{ccccc}
      $\mu a$ & \# config. & $P(\nu=0)$ & $P(\nu=1)$ & $P(\nu=2)$
      \\[1mm]\hline\\[-3mm]
      0.0 & 6783 & 0.544 & 0.426 & 0.029\\
      0.1 & 8703 & 0.526 & 0.439 & 0.034\\
      0.2 & 5760 & 0.500 & 0.454 & 0.046\\
      0.3 & 5760 & 0.476 & 0.465 & 0.058\\
      1.0 & 2816 & 0.268 & 0.446 & 0.210
    \end{tabular}
  \end{ruledtabular}
\end{table}

\begin{figure*}[-t]
  \centerline{\includegraphics[height=91mm]{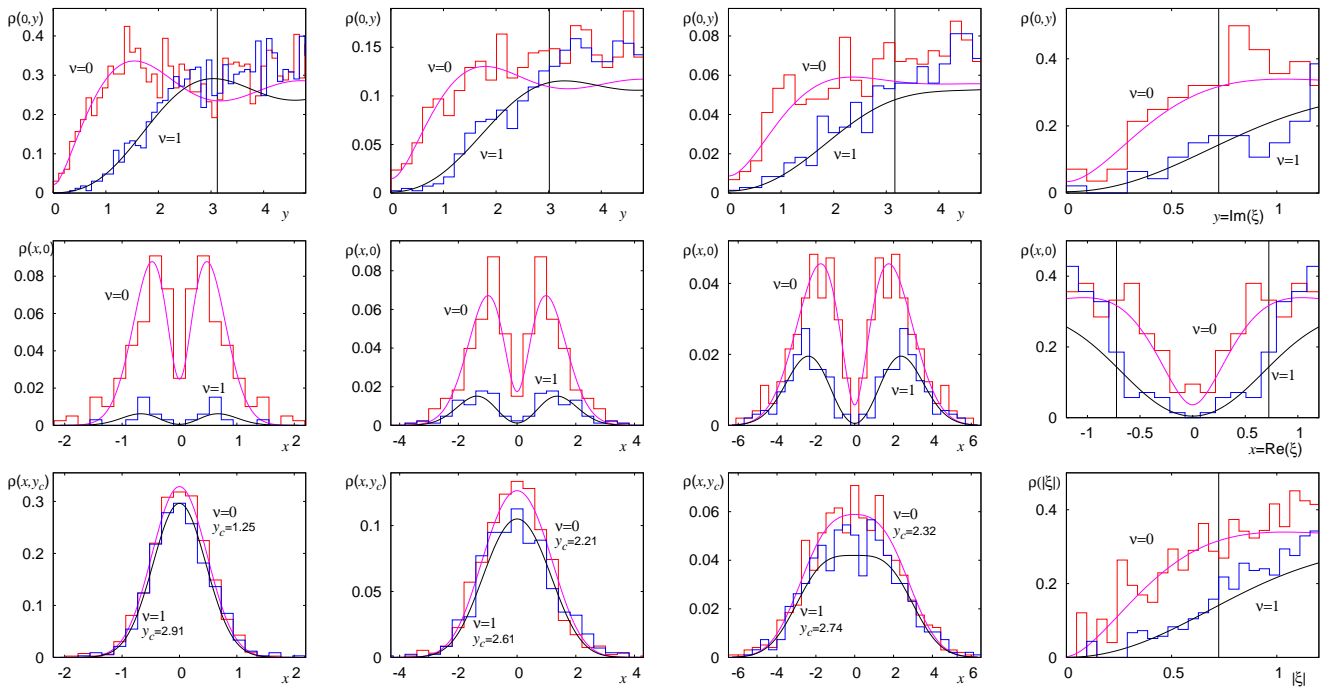}}
  \caption{Density of the small eigenvalues of $D_\text{ov}(\mu)$ in
    the complex plane (after projecting
    $\lambda\to2\lambda/(2-\lambda)$ and rescaling by $V\Sigma$) for
    (from left to right) $\mu=0.1,0.2,0.3,1.0$.  The histograms are
    lattice data for $\nu=0$ and $\nu=1$, and the solid lines are the
    corresponding RMT prediction of Eq.~\eqref{eq:rmt}, integrated
    over the bin size.  Top: cut along the imaginary axis, middle: cut
    along the real axis, bottom: cut parallel to the real axis.  The
    vertical lines indicate the fit interval.  For $\mu=1.0$, the
    eigenvalues are rescaled by $V\Sigma/2\sqrt\alpha$, the fit is a
    one-parameter fit to Eq.~\eqref{eq:strong}, and in the bottom plot
    the data are integrated over the phase.}
  \label{fig:rmt1}
\end{figure*}

Our parameters are given in Table~\ref{table1}.  In
Fig.~\ref{fig:spectrum} we show the spectrum of $D_\text{ov}(\mu)$ for
a typical configuration for $\mu=0$ and $\mu=0.3$.  As expected, we
see that the eigenvalues move away from the circle as $\mu$ is turned
on.  Another observation is that the number of zero modes of
$D_\text{ov}(\mu)$ for a given configuration can change as a function
of $\mu$, see Table~\ref{table1}.  This can be understood from the
relation between the anomaly and the index of $D_\text{ov}$
\cite{anomaly1,anomaly2},
\begin{align}
  \label{eq:index}
  -\tr(\gamma_5D_\text{ov})=2\ind(D_\text{ov})\:,
\end{align}
which we can show to remain valid at $\mu\ne0$.  Using
$\tr(\gamma_5D_\text{ov})=\tr[\gamma_5+\eps(\gamma_5D_W)]
=\tr[\eps(\gamma_5D_W)]$ and the fact that the eigenvalues of the sign
function are $+1$ or $-1$, one has $n_-^W-n_+^W=2\ind(D_\text{ov})$,
where $n_\pm^W$ denotes the number of eigenvalues of
$\gamma_5D_W(\mu)$ with real part $\gtrless0$.  As $\mu$ changes, an
eigenvalue of $\gamma_5D_W$ can move across the imaginary axis.  As a
result, $n_-^W-n_+^W$ changes by 2, and thus $\ind(D_\text{ov})$
changes by 1, which explains the observation.  We believe that this is
a lattice artefact which will disappear in the continuum limit.
(Also, the index theorem is violated at $\beta=5.1$, but for the
comparison with RMT only the number of zero modes matters.)

The spectral density of $D_\text{ov}(\mu)$ is given by
$\rho^\text{ov}(\lambda_r,\lambda_i)=\langle\sum_k
\delta(\lambda-\lambda_k)\rangle$ with $\lambda=\lambda_r+i\lambda_i$,
where the average is over configurations.  The claim is that the
distribution of the small eigenvalues of $D_\text{ov}(\mu)$ is
universal and given by RMT.  The RMT model for the Dirac operator is
\cite{Step96}
\begin{align}
  \label{eq:rmt-model}
  D_\text{RMT}(\mu)=
  \begin{pmatrix}0&iW+\mu\\iW^\dagger+\mu&0\end{pmatrix}\:,
\end{align}
where $W$ is a complex matrix of dimension $n\times(n+\nu)$ with no
further symmetries (we take $\nu\ge0$ without loss of generality).
The matrix in Eq.~\eqref{eq:rmt-model} has $\nu$ eigenvalues equal
to zero.  The spectral correlations of $D_\text{RMT}(\mu)$ on the
scale of the mean level spacing were computed in
Refs.~\cite{sv04,o04,aosv05}.  In the quenched approximation, the
result for the microscopic spectral density reads
\begin{align}
  \label{eq:rmt}
  \rho_s^\text{RMT}(x,y)&=\frac {x^2+y^2}{2\pi\alpha}
  e^{\frac{y^2-x^2}{4\alpha}} 
  K_\nu\left(\frac {x^2+y^2}{4\alpha}\right)\notag\\
  &\quad\times\int_0^1 t\, dt\, e^{-2\alpha t^2}|I_\nu(t z)|^2\:,
\end{align}
where $z=x+iy$, $I$ and $K$ are modified Bessel functions, and
$\alpha=\mu^2 f_\pi^2 V$.  To compare the lattice data to this result,
the eigenvalues $\lambda_k$ need to be rescaled by a parameter
$1/V\Sigma$ which is proportional to the mean level spacing near zero,
\begin{align}
  \label{eq:lat}
  \rho_s^\text{ov}(x,y)=\frac1{(V\Sigma)^2} \rho^\text{ov}\left(\frac
    x{V\Sigma},\frac y{V\Sigma}\right)\:.
\end{align}
$\Sigma$ and $f_\pi$ are low-energy constants that can be obtained
from a two-parameter fit of the lattice data, Eq.~\eqref{eq:lat}, to
the RMT prediction, Eq.~\eqref{eq:rmt}.  (The normalization is fixed
by $\int dx\,dy\,\rho^\text{ov}(x,y)=12V$ and does not introduce
another parameter.)  For $x\ll\alpha$, Eq.~\eqref{eq:rmt} simplifies
to \cite{jac-lectures}
\begin{align}
  \label{eq:strong}
  \rho_s^\text{RMT}(x,y)\to\frac{\xi}{2\pi\alpha}
  K_\nu\left(\xi\right)I_\nu\left(\xi\right)
  \text{ with } \xi=\frac{|z|^2}{4\alpha}\:.
\end{align}
A fit of Eq.~\eqref{eq:lat} to Eq.~\eqref{eq:strong} then only
involves the single parameter $\Sigma/f_\pi$.

In Fig.~\ref{fig:rmt1} we compare our lattice data to the RMT
prediction.  We display various cuts of the eigenvalue density in the
complex plane as explained in the figure captions.  The data agree
with the RMT predictions within our statistics.  $\Sigma$ and $f_\pi$
were obtained by a combined fit to the $\nu=0$ and $\nu=1$ data for
all three cuts and are displayed in Table~\ref{table2}.  (These
numbers have no physical significance at $\beta=5.1$.)  Separate fits
to the $\nu=0$ and $\nu=1$ data give results consistent with those in
Table~\ref{table2}.  For $\mu=1.0$ we have fitted to
Eq.~\eqref{eq:strong} since the distribution of the small eigenvalues
is radially symmetric up to $|\xi|\sim0.7$, and therefore $\Sigma$ and
$f_\pi$ cannot be determined independently by a fit to
Eq.~\eqref{eq:rmt}.  We note in passing that unfolding along the
imaginary axis results in better agreement of the lattice data with
RMT for larger values of $y$, but we shall not discuss this issue here
since it is a finite-volume effect that will be unimportant on larger
lattices.

\begin{table}[-t]
  \caption{Fit results for $\Sigma$ and $f_\pi$ (see text).}
  \label{table2}
  \begin{ruledtabular}
    \begin{tabular}{clclc}
      $\mu a$ & \multicolumn{1}{c}{$\Sigma a^3$} & $f_\pi a$ &
      \multicolumn{1}{c}{$\Sigma a^3/f_\pi a$} &
      \multicolumn{1}{c}{$\chi^2/\text{dof}$}\\[1mm]\hline \\[-3mm]
      0.0 & 0.0816(6) & \multicolumn{1}{c}{--} &
      \multicolumn{1}{c}{--} & 1.10 \\ 
      0.1 & 0.0812(11) & 0.261(6) & 0.311(5) & 0.67\\
      0.2 & 0.0785(14) & 0.245(5) & 0.320(4) & 0.78\\
      0.3 & 0.0824(17) & 0.248(5) & 0.332(4) & 1.03\\
      1.0 & \multicolumn{1}{c}{--} & \multicolumn{1}{c}{--} &
      0.603(18) & 0.42 
    \end{tabular}
  \end{ruledtabular}
\end{table}

The RMT results are valid in the ``ergodic regime''
\begin{align}
  \label{eq:ergodic}
  m_\pi,\mu\ll\frac1L\ll \Lambda\:,
\end{align}
where $L^4=V$, $m_\pi$ is the Goldstone boson mass, and $\Lambda$ is
the mass scale of the lightest non-Goldstone particle.  The rescaled
eigenvalues $z=\lambda V\Sigma$ should therefore be described by RMT
for $|z|\ll f_\pi^2\sqrt{V}$ (the dimensionless Thouless energy).  For
our data, $f_\pi^2\sqrt{V}\approx 1$ (this explains the choice of
$\beta=5.1$; for larger $\beta$ this number would be even smaller).
Nevertheless, the data seem to be described by RMT quite well even a
bit beyond this expectation.  Also, $\mu=1.0$ corresponds to $\mu
L=4$, which violates the inequality \eqref{eq:ergodic}.  The fact that
RMT still works below the Thouless energy in this case means that the
zero-momentum modes decouple from the partition function.  A detailed
study of the range of validity of RMT at $\mu\ne0$ as a function of
the lattice parameters will be the subject of further work (see also
Ref.~\cite{OW05}).

In summary, we have shown how to include a quark chemical potential in
the overlap operator.  The operator still satisfies a Ginsparg-Wilson
relation and has exact zero modes.  The distribution of its small
eigenvalues agrees with predictions of nonhermitian RMT for trivial
and nontrivial topology.  Our initial lattice study should be extended
to weaker coupling, larger lattices, and better statistics.  Work on
approximative methods to enable such studies is in progress.  For
small volumes, reweighting with the fermion determinant should allow
us to test RMT predictions for the unquenched theory \cite{AB06}.

\begin{acknowledgments}
  This work was supported in part by DFG.  We acknowledge illuminating
  discussions with W. Bietenholz, F.~Knechtli, H.~Neuberger, and
  J.J.M.~Verbaarschot (whom we also thank for a critical reading of
  the manuscript).  The simulations were performed on a \mbox{QCDOC}
  machine in Regensburg using USQCD software and Chroma \cite{usqcd}.
  We thank K.~Goto for providing us with an optimized GotoBLAS for
  QCDOC.
\end{acknowledgments}

\end{document}